\documentclass[12pt]{article}

\usepackage{amsmath}
\usepackage{amssymb}
\usepackage{amsfonts}
\usepackage{latexsym}
\usepackage{color}

\catcode `\@=11 \@addtoreset{equation}{section}
\def\theequation{\arabic{section}.\arabic{equation}}
\catcode `\@=12

\newcommand{\be}{\begin{equation}}
\newcommand{\en}{\end{equation}}
\newcommand{\bea}{\begin{eqnarray}}
\newcommand{\ena}{\end{eqnarray}}
\newcommand{\beano}{\begin{eqnarray*}}
\newcommand{\enano}{\end{eqnarray*}}
\newcommand{\bee}{\begin{enumerate}}
\newcommand{\ene}{\end{enumerate}}

\newcommand{\N}{\mathfrak N}

\newcommand{\mc}{\mathcal}

\newcommand{\E}{{\cal E}}
\newcommand{\F}{{\cal F}}

\newcommand{\Lc}{{\cal L}}

\newcommand{\1}{1 \!\! 1}

\newcommand{\Hil}{\mc H}

\newcommand{\ip}[2]{\left\langle {#1},{#2}\right\rangle}
\newcommand{\ipp}[2]{\left\langle {#1},{#2}\right\rangle_\pi}

\catcode `\@=11 \@addtoreset{equation}{section}
\catcode `\@=12

\begin{document}

\thispagestyle{empty}

\vspace*{2cm}

\begin{center}
{\Large \bf  Pseudo-bosons for the $D_2$ type quantum Calogero model}\\[10mm]

{\large F. Bagarello} \\
{ DEIM,
Facolt\`a di Ingegneria,\\ Universit\`a di Palermo,\\ I-90128  Palermo, ITALY\\
e-mail: fabio.bagarello@unipa.it\\
 Home page:
www.unipa.it/fabio.bagarello}
\vspace{3mm}\\

\end{center}

\vspace*{2cm}

\begin{abstract}
\noindent In the first part of this paper we show how a simple system, a 2-dimensional quantum harmonic oscillator, can be described in
terms of pseudo-bosonic variables. This apparently {\em strange} choice is useful when the {\em natural} Hilbert space of the system,
$\Lc^2({\Bbb R}^2)$ in this case, is, for some reason, not the most appropriate. This is exactly what happens for the $D_2$ type quantum Calogero model
considered in the second part of the paper, where the Hilbert space $\Lc^2({\Bbb R}^N)$ appears to be an unappropriate choice, since the
eigenvectors of the relevant hamiltonian are not square-integrable. Then we discuss how a certain intertwining operator arising from the model
can be used to fix a different Hilbert space more {\em useful}.

\end{abstract}

\vspace{2cm}


\vfill


\newpage

\section{Introduction}

In a series of recent papers we have investigated some mathematical and physical aspects of the so-called {\em pseudo-bosons}, see
\cite{bagrep} for a recent review. They arise from the canonical commutation relation $[a,a^\dagger]=\1$ upon replacing $a^\dagger$ by another
(unbounded) operator $b$ not (in general) related to $a$: $[a,b]=\1$. We have shown that $N=ba$ and $N^\dagger=a^\dagger b^\dagger$ can be both
diagonalized, and that their eigenvalues coincide with the set of natural numbers (including 0), ${\Bbb N}_0$. However the sets of related
eigenvectors are not orthonormal (o.n.) bases, while they turn out to be automatically {\em biorthogonal\/}. In most of the concrete examples
considered so far, they are bases of the Hilbert space of the system, $\Hil$, and, in some cases, they are also {\em Riesz bases\/}. In
\cite{bagnlpb} our original framework has been extended in order to include in the treatment also non self-adjoint operators with {\em more
elaborate} spectra, i.e. with eigenvalues not necessarily coincident with the natural numbers, while in \cite{bit2011} we have considered the
same problem from an algebraic point of view, constructing, in particular, the partial *-algebra arising from $a$ and $b$.

In a recent paper, \cite{ban}, the authors have shown how the $D_N$ type quantum Calogero model can be mapped, via a similarity transformation,
into a system of $N$ decoupled quantum harmonic oscillators (QHOs). However, in order to deduce the eigenvectors of the original hamiltonian, a
serious problem appears: they do not belong to $\Lc^2({\Bbb R}^N)$. For this reason, they propose a possible way out by considering suitable
symmetrized wave-functions and showing that these are o.n. with respect to a different scalar product. Similar problems already appeared in
older papers, regarding other types of Calogero models ($A_{N-1}$ and $B_N$, for instance, in \cite{nishino} and in references therein).

In this paper, we will consider a similar approach, but we justify our proposal by a preliminary analysis of a two-dimensional QHO. This is
related to the possibility of being interested to functions, exactly as  it happens for the Calogero model, which do not belong to $\Lc^2({\Bbb
R}^2)$ but to some different Hilbert space. Moreover we show that, with this choice, we end up with a simple o.n. basis and that the
hamiltonian $\tilde H_D$, see Section IV, which is not self-adjoint in $\Lc^2({\Bbb R}^2)$, becomes self-adjoint (at least formally) in the
{\em new} Hilbert space.

This paper is organized as follows:  in the next section we briefly review few facts on  two-dimensional pseudo-bosons.  In Section III we show
how they can be used, and why, in the analysis of a two (or more)-dimensional QHO. In Section IV we write the $D_2$ type quantum Calogero model
in terms of pseudo-bosonic operators and we discuss the problems arising and a possible solution. Our conclusions are given in Section V. The
appendix contains some technical results.

\section{The commutation rules}

Let $\Hil$ be a given Hilbert space with scalar product $\left<.,.\right>$ and related norm $\|.\|$. We introduce two pairs of operators, $a_j$
and $b_j$, $j=1,2$, acting on $\Hil$ and satisfying the following commutation rules \be [a_j,b_j]=\1, \quad \mbox{ and }\quad
[a_1^\sharp,a_2^\sharp]=[a_1^\sharp,b_2^\sharp]=[b_1^\sharp,b_2^\sharp]=0, \label{21} \en where $x^\sharp$ stands for $x$ or $x^\dagger$
($x=a_j, b_j$). Of course, they collapse to the CCR's for independent modes if $b_j=a^\dagger_j$, $j=1,2$. It is well known that $a_j$ and
$b_j$  cannot be defined on all of $\Hil$. Writing $D^\infty(X):=\cap_{p\geq0}D(X^p)$ (the common  domain of all the powers of the operator
$X$), we consider the following:

\vspace{2mm}

{\bf Assumption 1.--} there exists a non-zero $\varphi_{0,0}\in\Hil$ such that $a_j\varphi_{0,0}=0$, $j=1,2$, and $\varphi_{0,0}\in
D^\infty(b_1)\cap D^\infty(b_2)$.

{\bf Assumption 2.--} there exists a non-zero $\Psi_{0,0}\in\Hil$ such that $b_j^\dagger\Psi_{0,0}=0$, $j=1,2$, and $\Psi_{0,0}\in
D^\infty(a_1^\dagger)\cap D^\infty(a_2^\dagger)$.

\vspace{2mm}

Under these assumptions we can introduce the following vectors in $\Hil$: \be \varphi_{n,l}=\frac{1}{\sqrt{n!l!}}\,b_1^n\,b_2^l\,\varphi_{0,0}
\quad \mbox{ and }\quad \Psi_{n,l}=\frac{1}{\sqrt{n!l!}}(a_1^\dagger)^n(a_2^\dagger)^l\Psi_{0,0}, \quad n,l\geq 0. \label{22}\en Let us now
define the unbounded operators $N_j:=b_ja_j$ and $\N_j:=N_j^\dagger=a_j^\dagger b_j^\dagger$, $j=1,2$.  It is possible to check that
$\varphi_{n,l}$ belongs to the domain of $N_j$, $D(N_j)$, and that $\Psi_{n,l}\in D(\N_j)$, for all $n,l\geq0$ and $j=1,2$. Moreover, \be
N_1\varphi_{n,l}=n\varphi_{n,l}, \quad N_2\varphi_{n,l}=l\varphi_{n,l}, \quad \N_1\Psi_{n,l}=n\Psi_{n,l}, \quad \N_2\Psi_{n,l}=l\Psi_{n,l}.
\label{23}\en

Under the above assumptions, and choosing the normalization of $\Psi_{0,0}$ and $\varphi_{0,0}$ in such a way
$\left<\Psi_{0,0},\varphi_{0,0}\right>=1$, it is  easy to check that
 \be
\left<\Psi_{n,l},\varphi_{m,k}\right>=\delta_{n,m}\delta_{l,k}, \quad \forall n,m,l,k\geq0. \label{27}\en This means that the sets
$\F_\Psi=\{\Psi_{n,l},\,n,l\geq0\}$ and $\F_\varphi=\{\varphi_{n,l},\,n,l\geq0\}$ are {\em biorthogonal} and, therefore, the vectors of
each set are linearly independent. However, they are not necessarily complete in $\Hil$. Hence, this is something to be assumed:  \vspace{2mm}

{\bf Assumption 3.--} $\F_\Psi$ and $\F_\varphi$ are bases of $\Hil$.

\vspace{2mm}

Let us now introduce the operators $S_\varphi$ and $S_\Psi$ via their action respectively on  $\F_\Psi$ and $\F_\varphi$: \be
S_\varphi\Psi_{n,k}=\varphi_{n,k},\qquad S_\Psi\varphi_{n,k}=\Psi_{n,k}, \label{213}\en for all $n, k\geq0$, which also imply that
$\Psi_{n,k}=(S_\Psi\,S_\varphi)\Psi_{n,k}$ and $\varphi_{n,k}=(S_\varphi \,S_\Psi)\varphi_{n,k}$, for all $n,k\geq0$. Hence, at least if
$S_\varphi$ and $S_\Psi$ are bounded,  \be S_\Psi\,S_\varphi=S_\varphi\,S_\Psi=\1 \quad \Rightarrow \quad S_\Psi=S_\varphi^{-1}. \label{214}\en
In other words, both $S_\Psi$ and $S_\varphi$ are invertible and one is the inverse of the other. Furthermore, we can also check that they are
both positive, well defined and symmetric, and they can be written, formally, in the bra-ket notation as \be S_\varphi=\sum_{n,k=0}^\infty
|\varphi_{n,k}><\varphi_{n,k}|,\qquad S_\Psi=\sum_{n,k=0}^\infty |\Psi_{n,k}><\Psi_{n,k}|. \label{212}\en
 These operators are both bounded if and only if $\F_\Psi$ and $\F_\varphi$ are Riesz bases, \cite{bagpb1}. In this case the pseudo-bosons
 have been called {\em regular}.
 We conclude this short review by recalling that
the following interesting intertwining relations among non self-adjoint operators, are automatically satisfied: \be S_\Psi\,N_j=\N_jS_\Psi \quad \mbox{ and }\quad
N_j\,S_\varphi=S_\varphi\,\N_j, \label{219}\en $j=1,2$. These are related to the fact that: (i) $N_j$ and $N_j^\dagger$ have the same
eigenvalues; (ii) their eigenvectors are related by $S_\Psi$ and $S_\varphi$ as in (\ref{213}).

\section{The QHO}

In view of what we will do in the next sections, we consider here  a multi-dimensional QHO\footnote{As a matter of fact, we will only consider
a two-dimensional case, since adding dimensions to the system does not change significantly our main results.} in terms of pseudo-bosonic
operators. Of course, in principle, there is no reason a priori to replace ordinary canonical commutation relations with something like
(\ref{21}). However, it may happen, and this is exactly what happens in Section IV, that for some reason we need to act on some particular
square integrable functions with unbounded operators which {\em bring} these functions outside of $\Hil=\Lc^2({\Bbb R}^2)$. This suggests that
$\Lc^2({\Bbb R}^2)$ may not be the {\em right} Hilbert space. For this reason we consider the following question: what does it happen if we
consider the two-dimensional QHO in a Hilbert space $\Hil_\pi$ which is not necessarily $\Lc^2({\Bbb R}^2)$? In particular, we will consider
here $\Hil_\pi=\Lc^2({\Bbb R}^2,e^{\pi(x_1,x_2)}dx_1\,dx_2)$, where $\pi(x_1,x_2)\leq0$ almost everywhere (a.e.) is a function still to be
fixed. The scalar product in $\Hil_\pi$ is
$$
\ipp{f}{g}:=\int\int_{{\Bbb R}^2}\,\overline{f(x_1,x_2)}\,g(x_1,x_2)\,e^{\pi(x_1,x_2)}dx_1\,dx_2,
$$
for all $f, g$ for which this integral makes sense. For instance, later on we will take $\pi(x_1,x_2)=-\,\frac{1}{2}\,\omega(x_1^2+x_2^2)$, for
a given positive $\omega$. With this choice, as well as for any other choice of non positive $\pi(x_1,x_2)$, it is clear that $\Hil_\pi\supset\Hil$, so that many functions which are not square-integrable,
can still be treated as elements of  $\Hil_\pi$. This is the situation we are interested in, here. The price we have to
pay, how we will discuss in details, is that the orthonormality of the standard set of eigenvectors of the hamiltonian of the oscillator, $h$,
is lost, as well as its self-adjointness. This is due to the fact that in $\Hil_\pi$ the adjoint, $*$, does not coincide with that of
$\Hil$, $\dagger$. In fact, calling $\ip{.}{.}$ the scalar product in $\Hil$, they satisfy different rules:
$$
\ip{Af}{g}=\ip{f}{A^\dagger g}, \quad \mbox{ and }\quad \ipp{Af}{g}=\ipp{f}{A^* g},
$$
whenever these quantities are defined. Then, the position and the momentum operators satisfy the following: $x_j^\dagger=x_j=x_j^*$, while
$p_j=p_j^\dagger$ but $p_j^*=p_j-i\pi_j(x_1,x_2)$, $j=1,2$, where $\pi_j(x_1,x_2)=\frac{\partial\pi(x_1,x_2)}{\partial x_j}$. Of course, if, in
particular, $\pi(x_1,x_2)=0$ a.e., then $\Hil_\pi=\Hil$ and $\dagger=*$.

The hamiltonian of our QHO is $h=\frac{1}{2}(-\partial_1^2+\omega^2x_1^2)+\frac{1}{2}(-\partial_2^2+\omega^2x_2^2)$. This hamiltonian is self
adjoint in $\Hil$, $h=h^\dagger$, while it is not self-adjoint in $\Hil_\pi$, $h\neq h^*$. Introducing, as usual,
$a_j=\frac{1}{\sqrt{2\omega}}(\omega x_j+ip_j)=\frac{1}{\sqrt{2\omega}}(\omega x_j+\partial_j)$ and
$b_j=a_j^\dagger=\frac{1}{\sqrt{2\omega}}(\omega x_j-\partial_j)\neq a_j^*$, we can rewrite $h$, but for an unessential constant, as follows
$$
h=\omega\left(a_1^\dagger\,a_1+a_2^\dagger\,a_2\right)=\omega\left(b_1\,a_1+b_2\,a_2\right),
$$
depending on the Hilbert space we want to use for the analysis of the system. In this section we use $h=\omega\left(b_1\,a_1+b_2\,a_2\right)$
since, as we have already stressed, we are interested in working in $\Hil_\pi$.

It is clear that the function $\varphi_{0,0}(x_1,x_2)$ satisfying Assumption 1 of Section II is, except for a normalization, exactly the {\em
standard gaussian}: $\varphi_{0,0}(x_1,x_2)=N_\varphi\,e^{-\frac{1}{2}\omega(x_1^2+x_2^2)}$. In fact,
$a_1\varphi_{0,0}(x_1,x_2)=a_2\varphi_{0,0}(x_1,x_2)=0$ and $b_1^{n_1}b_2^{n_2}\varphi_{0,0}(x_1,x_2)$ are well defined functions in
$\Hil\subset\Hil_\pi$ for all $n_1, n_2\geq0$. This follows from the fact that $b_j=a_j^\dagger$, so that
\be\varphi_{n_1,n_2}(x_1,x_2)=\frac{N_\varphi}{\sqrt{n_1!n_2!2^{n_1+n_2}}}\,H_{n_1}(\sqrt{\omega}x_1)H_{n_2}(\sqrt{\omega}x_2)
e^{-\frac{1}{2}\omega(x_1^2+x_2^2)},\label{extra}\en where $H_n$ is the $n$-th Hermite polynomial. For all $n_1$ and $n_2$ these functions
belong to $\Hil$, so that they also belong to $\Hil_\pi$.

The set $\F_\varphi$ of these functions is a basis in $\Hil_\pi$: it is clear that the various $\varphi_{n_1,n_2}$ are linearly independent. We
also know that $\F_\varphi$ is complete in $\Hil$, and is actually an o.n. basis for $\Hil$, for a certain choice of $N_\varphi$. In order to
check that $\F_\varphi$ is complete in $\Hil_\pi$ too, we consider $f\in\Hil_\pi$ satisfying $\ipp{f}{\varphi_{n_1,n_2}}=0$, $\forall\,n_1,
n_2$. Now, since $\pi(x_1,x_2)\leq0$ a.e., $e^{2\pi(x_1,x_2)}\leq e^{\pi(x_1,x_2)}$ a.e., so that $fe^\pi\in\Hil$. In fact
$$
\|f\,e^\pi\|^2=\int\int_{{\Bbb R}^2}\,|f(x_1,x_2)|^2\,e^{2\pi(x_1,x_2)}dx_1\,dx_2\leq \int\int_{{\Bbb R}^2}\,|f(x_1,x_2)|^2\,e^{\pi(x_1,x_2)}dx_1\,dx_2=\|f\|^2_\pi,
$$
which is finite since $f\in\Hil_\pi$. Hence, we can write $0=\ipp{f}{\varphi_{n_1,n_2}}=\ip{f\,e^{\pi}}{\varphi_{n_1,n_2}}$, $\forall\,n_1,
n_2$. The completeness of $\F_\varphi$ in $\Hil$ implies now that $f\,e^\pi=0$. Therefore $f=0$.

Just a little bit more complicated is the analysis of Assumption 2, since the role of $\pi(x_1,x_2)$ appears already  in the definition of the functions
$\psi_{n_1,n_2}(x_1,x_2)$, and not only at the level of the scalar product. Indeed we have
$$
b_j^*=\frac{1}{\sqrt{2\omega}}(\omega x_j+\pi_j+\partial_j),\qquad a_j^*=\frac{1}{\sqrt{2\omega}}(\omega x_j-\pi_j-\partial_j).
$$
Notice that, not surprisingly, $b_j^*=(a_j^\dagger)^*\neq a_j$. The vacuum of $b_j^*$ can be easily computed:
$\Psi_{0,0}(x_1,x_2)=N_\Psi\,e^{-\frac{1}{2}\omega(x_1^2+x_2^2)-\pi(x_1,x_2)}$. A sufficient condition for $\Psi_{0,0}$ to belong to $\Hil_\pi$
is that $\pi(x_1,x_2)=(\beta-\omega)(x_1^2+x_2^2)$, with $0<\beta<\omega$. With this choice, for instance,
$\|\Psi_{0,0}\|_\pi^2=|N_\Psi|^2\,\frac{\pi}{\omega}$. Moreover, the action of powers of $a_j^*$ on $\Psi_{0,0}(x_1,x_2)$ produces polynomials
times an exponential function, not necessarily decreasing,
$$
\Psi_{n_1,n_2}(x_1,x_2)=N_\Psi\,p_{n_1}(x_1)q_{n_2}(x_2)e^{\left(\frac{\omega}{2}-\beta\right)(x_1^2+x_2^2)},
$$
which are still in $\Hil_\pi$ for all $n_1$ and $n_2$, due to the definition of $\ipp{.}{.}$. Of course, if $\pi(x_1,x_2)=\pi(x_2,x_1)$, for all $n$ we have $p_n(x)=q_n(x)$, where $n$ is the order of the polynomials. Otherwise these polynomials may be different. Hence,
Assumption 2 is satisfied. In Section \ref{sec3a}, with the particular choice of $\beta=\frac{\omega}{2}$, we will give the details of these
polynomials. The functions $\Psi_{n_1,n_2}$ are linearly independent, clearly. To check that $\F_\Psi=\{\Psi_{n_1,n_2}\}$ is also complete in
$\Hil_\pi$, we begin noticing that $f\in\Hil_\pi$ if and only if $f\,e^{\frac{1}{2}\pi}\in\Hil$: $\|f\|_\pi=\|f\,e^{\frac{1}{2}\pi}\|$.
Secondly, if $p_n$  and $q_m$ are polynomials  of degree $n$ and $m$ respectively, it is possible to check, \cite{kolfom}, that the auxiliary
set $\F_p=\{P_{n_1,n_2}(x_1,x_2):=p_{n_1}(x_1)q_{n_2}(x_2)e^{-\frac{\beta}{2}(x_1^2+x_2^2)}, n_j\geq0\}$, is complete in $\Lc^2({\Bbb R}^2)$
for all choices of positive $\beta$. Therefore, assuming that  $\ipp{f}{\Psi_{n_1,n_2}}=0$ for all $n_1$ and $n_2$, since
$$
\ipp{f}{\Psi_{n_1,n_2}}=\int\int_{{\Bbb R}^2}\left(\overline{f(x_1,x_2)}\,e^{\frac{1}{2}\,\pi(x_1,x_2)}\right)\,P_{n_1,n_2}(x_1,x_2)\,dx_1\,dx_2,
$$
we deduce that $f(x_1,x_2)\,e^{\frac{1}{2}\,\pi(x_1,x_2)}=0$ a.e., so that $f(x_1,x_2)=0$ a.e., as expected. Hence, both $\F_\varphi$ and
$\F_\Psi$ are complete in $\Hil_\pi$.

\subsection{The choice $\pi(x_1,x_2)=-\frac{\omega}{2}(x_1^2+x_2^2)$}
\label{sec3a}

Let us now consider what happens if $\beta=\frac{\omega}{2}$ or, in other words, when $\pi(x_1,x_2)=-\frac{\omega}{2}(x_1^2+x_2^2)$. Of course,
nothing changes in $\F_\varphi$, since $\pi(x_1,x_2)$ plays no role in the definition of $\varphi_{n_1,n_2}(x_1,x_2)$. On the other hand, since
$a_j^*=\frac{1}{\sqrt{2\omega}}\left(2\omega\,x_j-\partial_j\right)$ and $b_j^*=\frac{1}{\sqrt{2\omega}}\partial_j$, we find \be
\Psi_{n_1,n_2}(x_1,x_2)=\frac{N_\Psi}{\sqrt{n_1!n_2!2^{n_1+n_2}}}\,H_{n_1}(\sqrt{\omega}\,x_1)H_{n_2}(\sqrt{\omega}\,x_2), \label{extra2}\en
where $H_n(x)$ is the $n$-th Hermite polynomial. Incidentally we observe that $\Psi_{n_1,n_2}(x_1,x_2)\notin\Hil$, while
$\Psi_{n_1,n_2}(x_1,x_2)\in\Hil_\pi$. This remark will be our main motivation for changing Hilbert space in the analysis of the $D_2$ type of
the Calogero model in the next section. Fixing the normalization $N_\varphi=N_\Psi=\sqrt{\frac{\omega}{\pi}}$ we get
$\ipp{\varphi_{n_1,n_2}}{\Psi_{m_1,m_2}}=\delta_{n_1,m_1}\delta_{n_2,m_2}$: the two sets are biorthogonal. They are more than this: they are
bases. The proof of this statement is given in the Appendix.

Using now the summation formula $\sum_{k=0}^\infty\,\frac{1}{2^kk!}\,H_k(x)\,H_k(y)=\sqrt{\pi}\,e^{\frac{1}{2}(x^2+y^2)}\delta(x-y)$, we deduce
the following forms for the intertwining operators $S_\varphi$ and $S_\Psi$:
$$
\left(S_\Psi f\right)(x_1,x_2):=\sum_{n_1,n_2=0}^\infty\ipp{\Psi_{n_1,n_2}}{f}\Psi_{n_1,n_2}(x_1,x_2)=e^{\frac{1}{2}\,\omega\,(x_1^2+x_2^2)}\,f(x_1,x_2),
$$
while
$$
\left(S_\varphi f\right)(x_1,x_2):=\sum_{n_1,n_2=0}^\infty\ipp{\varphi_{n_1,n_2}}{f}\varphi_{n_1,n_2}(x_1,x_2)=e^{-\frac{1}{2}\,\omega\,(x_1^2+x_2^2)}\,f(x_1,x_2),
$$
so that they are multiplication operators and $S_\varphi=S_\Psi^{-1}$, as expected. The operator $S_\Psi$ is unbounded on $\Hil_\pi$, while
$S_\varphi$ turns out to be bounded. Hence, see \cite{bagrep}, our pseudo-bosons are non-regular and, consequently, $\F_\varphi$ and $\F_\Psi$
are not Riesz bases.

\vspace{4mm}

Going back to the general situation, we have $h\varphi_{n_1,n_2}=\omega(n_1+n_2)\varphi_{n_1,n_2}$ and
$h^*\Psi_{n_1,n_2}=\omega(n_1+n_2)\Psi_{n_1,n_2}$, as well as $S_\Psi\,h=h^*S_\Psi$. Needless to say, all that we have found here can be
trivially extended to higher dimensions.

The lesson we learn is the following: if $\Lc^2({\Bbb R}^N)$ is not {\em big enough} to deal with a particular quantum system, we could always try
to introduce some weight in the Hilbert space (i.e. to change the scalar product) in order to {\em enlarge} it. The price to pay, which is not
an high price, is that the orthogonal set of eigenstates of a certain self-adjoint operator $h=h^\dagger$ must be doubled, producing two sets
of biorthogonal bases, which turn out to be eigenvectors of $h$ and of the new adjoint of, $h$, $h^*$, respectively. These two sets are further mapped one into the other by
a certain intertwining operator and by its inverse. Stated in different words: the use of pseudo-hermitian operators, \cite{mosta,bender}, could be related to the necessity of enlarge the Hilbert space where the physical system needs to be  described.

\section{The Calogero model in pseudo-bosonic variables}

We will now consider the $D_2$ type Calogero model as described in \cite{ban}. In particular, their preliminary analysis will play a crucial role in our treatment. For this reason, their approach will be quickly reviewed.

The starting point is the following self-adjoint (in $\Hil=\Lc^2({\Bbb R}^2)$) hamiltonian
$$H_D=\frac{1}{2}\sum_{j=1}^2\left(-\partial_j^2+\omega^2 x_j^2\right)+\nu(\nu-1)\left[\frac{1}{(x_1-x_2)^2}+\frac{1}{(x_1+x_2)^2}\right],$$
where $\nu>\frac{1}{2}$, which, putting $\Psi_0(x_1,x_2)=|x_1^2-x_2^2|^\nu\,e^{-\frac{1}{2}\omega(x_1^2+x_2^2)}$ and $E_0=\omega(1+2\nu)$,
produces \be\tilde H_D:=\Psi_0^{-1}(H_D-E_0)\Psi_0=\omega O_E-\frac{1}{2}\,O_L, \label{31}\en where we have introduced the operators
$O_E=x_1\partial_1+x_2\partial_2$, $\nabla^2=\partial_1^2+\partial_2$, $X^2=x_1^2+x_2^2$ and
$O_L=\nabla^2+4\nu\,\frac{1}{x_1^2-x_2^2}\left(x_1\partial_1-x_2\partial_2\right)$. They satisfy the following commutation relations \be
[O_L,O_E]=2O_L,\quad [O_E,X^2]=2X^2,\quad [\nabla^2,O_E]=2\nabla^2,\quad [\nabla^2,X^2]=4(O_E+\1). \label{32}\en Let us now put
$a_j=\frac{1}{\sqrt{2\omega}}\left(\omega x_j+ip_j\right)$, $a_j^\dagger=\frac{1}{\sqrt{2\omega}}\left(\omega x_j-ip_j\right)$, $j=1,2$, and
let us introduce, at least formally, the operator \be T=e^{-\frac{1}{4\omega}O_L}\,e^{\frac{1}{4\omega}\nabla^2}\,e^{\frac{1}{2}\omega X^2}.
\label{33}\en Here $a_j^\dagger$ is the adjoint of $a_j$ in $\Lc^2({\Bbb R}^2)$. The operator $T$ is invertible and we find \be
T^{-1}\tilde H_D T=\omega\hat N=:h, \label{34}\en where $\hat N=\hat n_1+\hat n_2=a_1^\dagger a_1+a_2^\dagger a_2$, \cite{ban}. It is interesting to notice
that, while $H_D$ and $h$ are self-adjoint, the intermediate hamiltonian $\tilde H_D$ is not. In particular we can check that $$\tilde
H_D^\dagger=-\tilde H_D-\nabla^2-4\nu\frac{x_1^2+x_2^2}{(x_1^2-x_2^2)^2}-2\omega\1. $$ This is related first to the fact that the map from $
H_D$ to $\tilde H_D$, implemented by $\Psi_0$, is not unitary. Also, $T^\dagger\neq T^{-1}$: $T$ is a similarity but not an unitary map. Hence,
as already discussed in \cite{ban}, domain problems arise, since $T$ and $T^{-1}$ are unbounded.

\vspace{2mm}

{\bf Remark:--} A similar procedure works also for larger values of $N$, \cite{ban}. In fact, introducing
$$
H_D=\frac{1}{2}\sum_{j=1}^N\left(-\partial_j^2+\omega^2 x_j^2\right)+\nu(\nu-1)\sum_{1\leq i< j\leq N}\left[\frac{1}{(x_i-x_j)^2}+\frac{1}{(x_i+x_j)^2}\right],
$$
$$
\Psi_0(x_1,x_2)=\prod_{1\leq i< j\leq N}|x_i^2-x_j^2|^\nu\,e^{-\frac{1}{2}\omega\sum_{i=1}^N x_i^2},\quad E_0=\frac{1}{2}N\omega+\nu N(N-1)\omega,
$$
and
$$
O_E=\sum_{i=1}^N x_i\partial_i,\quad O_L=\sum_{i=1}^N \partial_i^2+4\nu\,\sum_{1\leq i< j\leq N}\frac{1}{x_i^2-x_j^2}\left(x_i\partial_i-x_j\partial_j\right),
$$
we could repeat, with minor differences, the same steps as for $N=2$, mapping (non unitarily) $H_D$ into a self-adjoint hamiltonian describing $N$ uncoupled operators.

\subsection{A formal point of view in $\Hil$}

Let us introduce two pairs of operators $A_j:=Ta_jT^{-1}$ and $B_j=Ta_j^\dagger T^{-1}$, $j=1,2$. It is clear that $[A_j,B_k]=\delta_{j,k}\1$,
$A_j^\dagger\neq B_j$. Hence they are pseudo-bosonic operators, not everywhere defined. Moreover, it is possible to show that, introducing the
invertible but, again, not unitary operator $\Omega:=e^{-\frac{1}{4\omega}O_L}$, these operators take the following, simpler, forms: \be
A_j=\frac{i}{\sqrt{2\omega}}\,\Omega\, p_j \,\Omega^{-1},\qquad B_j=\sqrt{2\omega}\,\Omega\, x_j\, \Omega^{-1}. \label{35}\en These formulas
show that the pseudo-bosonic operators are (at least, formally) similar to the original position and momentum operators. The hamiltonian
$\tilde H_D$ can be written now as \be \tilde H_D=\omega(B_1A_1+B_2A_2). \label{36}\en The eigenstates of $\tilde H_D$ can be formally
constructed quite easily: let $\varphi_{0,0}$ be a non zero vector satisfying $A_j\varphi_{0,0}=0$, $j=1,2$, and let us define the vectors
$\varphi_{n_1,n_2}=\frac{1}{\sqrt{n_1!\,n_2!}}\,B_1^{n_1}\,B_2^{n_2}\varphi_{0,0}$, as in Section II. For the time being, we assume that all these vectors are
well defined in $\Hil$. We will return on this aspect in a moment. Then we have $\tilde H_D\varphi_{n_1,n_2}=\epsilon_{n_1,n_2}\varphi_{n_1,n_2}$,
where $\epsilon_{n_1,n_2}=\omega(n_1+n_2)$. Going on with our formal computations, it is quite easy to find the eigenstates of $\tilde
H_D^\dagger=\omega(A_1^\dagger\,B_1^\dagger+A_2^\dagger\,B_2^\dagger)$. For that it is enough to consider a second {\em ground state}
$\Psi_{0,0}$, satisfying $B_j^\dagger\Psi_{0,0}=0$, $j=1,2$, and to define the vectors
$\Psi_{n_1,n_2}=\frac{1}{\sqrt{n_1!\,n_2!}}\,(A_1^\dagger)^{n_1}\,(A_2^\dagger)^{n_2}\Psi_{0,0}$. Then $\tilde
H_D^\dagger\Psi_{n_1,n_2}=\epsilon_{n_1,n_2}\Psi_{n_1,n_2}$, and the $\varphi_{n_1,n_2}$'s and $\Psi_{n_1,n_2}$'s are expected to be mutually
orthogonal. Continuing our formal approach, we could further expect that $\F_\varphi=\{\varphi_{n_1,n_2}\}$ and $\F_\Psi=\{\Psi_{n_1,n_2}\}$ are
bases of $\Hil$, but not necessarily Riesz bases. $\F_\varphi$ and $\F_\Psi$ could be
further used to define two intertwining operators, as discussed in Section II.

However, the situation is not so simple, as already stressed in \cite{ban}. For this reason, we devote the next section to an alternative approach to the system.

\subsection{Back to non-formal computations}

\label{sectbtnfc}

Let us consider the hamiltonian $h$ introduced in (\ref{34}). This is the hamiltonian of a two-dimensional quantum oscillator, and its eigenstates are
easily found: these can be written as
$\Phi_{n_1,n_2}(x_1,x_2)=\frac{1}{\sqrt{n_1!\,n_2!}}\,(a_1^\dagger)^{n_1}\,(a_2^\dagger)^{n_2}\Phi_{0,0}(x_1,x_2)$, where
$a_j\Phi_{0,0}(x_1,x_2)=0$, $j=1,2$. Then $$h\,\Phi_{n_1,n_2}(x_1,x_2)=\epsilon_{n_1,n_2}\Phi_{n_1,n_2}(x_1,x_2).$$
 Explicitly we have
$$
\Phi_{n_1,n_2}(x_1,x_2)=\frac{1}{\sqrt{n_1!n_2!2^{n_1+n_2}}}\left(\frac{\omega}{\pi}\right)^{1/2}H_{n_1}(\sqrt{\omega}\,x_1)H_{n_2}(\sqrt{\omega}\,x_2)\,
e^{-\frac{1}{2}\,\omega(x_1^2+x_2^2)},
$$ where, as usual,  $H_n$ is the $n$-th Hermite polynomial.
It is now easy to write the functions in $\F_\varphi$ is a rather compact form: \be
\varphi_{n_1,n_2}(x_1,x_2)=T\Phi_{n_1,n_2}(x_1,x_2)=\left(\frac{\omega}{\pi}\right)^{1/2}\sqrt{\frac{(2\omega)^{n_1+n_2}}{n_1!n_2!}}\,\Omega\,
x_1^{n_1}\,x_2^{n_2}, \label{37}\en which reduces, for instance, to  $\varphi_{0,0}(x_1,x_2)=\left(\frac{\omega}{\pi}\right)^{1/2}$, for
$n_1=n_2=0$. This already shows that, contrarily to what assumed previously and in agreement with the results in \cite{ban,nishino} and references therein,
$\varphi_{0,0}(x_1,x_2)$ does not belong to $\Lc^2({\Bbb R}^2)$. It is natural to expect that, also for other values of $n_1$ and $n_2$,
$\varphi_{n_1,n_2}(x_1,x_2)\notin\Lc^2({\Bbb R}^2)$. Moreover, moving from formal to explicit computations, is even more difficult to deduce
the analytic form of the functions $\Psi_{n_1,n_2}(x_1,x_2)$, and to check whether they belong or not to $\Hil$. However, in Section III we
have shown in some details that leaving $\Lc^2({\Bbb R}^2)$ is not a major disaster, and can produce interesting consequences.  This is, in
fact, our point of view now, and we ask the following question: instead of working in $\Lc^2({\Bbb R}^2)$, does it exist some different Hilbert space which can be used to deal with
$\tilde H_D$ and with its eigenvectors? The answer is affirmative: the right Hilbert space is $\Hil_T:=D(T^{-1})$, the domain of the operator
$T^{-1}$,
endowed with the scalar product $\ip{f}{g}_T:=\ip{T^{-1}}{T^{-1}g}$, $f,g\in\Hil_T$\footnote{To be more rigorous, $T^{-1}$ should be a closed operator or, at least, closable. In this case, in fact,  we could use its closed extension to define the Hilbert space.}. It is clear that $\varphi_{n_1,n_2}\in\Hil_T$,
$\forall\,n_1, n_2$. More than this, the set $\F_\varphi$ is o.n.:
$$
\ip{\varphi_{n_1,n_2}}{\varphi_{m_1,m_2}}_T=\ip{T^{-1}(T\Phi_{n_1,n_2})}{T^{-1}(T\Phi_{m_1,m_2})}=\ip{\Phi_{n_1,n_2}}{\Phi_{m_1,m_2}}=\delta_{n_1,m_1}\delta_{n_2,m_2}.
$$
Of course, this implies that the set $\F_\Psi$ coincides with $\F_\varphi$ itself. This can be seen as a consequence of the uniqueness of the
biorthogonal basis for a given basis in a certain Hilbert space. The fact that $\varphi_{n_1,n_2}=\Psi_{n_1,n_2}$ for all $n_1$, $n_2$ may, at
a first sight, appear strange, since $\tilde H_D\neq\tilde H_D^\dagger$. However, this is not so since $\tilde H_D$ turns out to be self-adjoint in
$\Hil_T$. In fact, because of (\ref{34}),
$$
\ip{\tilde H_Df}{g}_T=\ip{T^{-1}\tilde H_Df}{T^{-1}g}=\ip{h\,T^{-1}f}{T^{-1}g}=\ip{T^{-1}f}{h\,T^{-1}g}=$$
$$=\ip{T^{-1}f}{T^{-1}\,\tilde H_D\,g}=\ip{f}{\tilde H_D\,g}_T,
$$
for all $f, g\in D(\tilde H_D)$. In a certain sense, this could be considered as a driving idea: whenever a non selfadjoint hamiltonian $\hat
H$ on $\Hil$ is similar to a self adjoint operator h, or when a (closed and injective) intertwining operator $X$ relating $\hat H$ and $h$ does
exist, $X\,\hat H=h\,X$, the {\em natural} Hilbert space where $\hat H$ {\em lives} is not $\Hil$ with its own scalar product, but rather than
that $\Hil_X:=D(X)$ with a scalar product $\ip{f}{g}_X:=\ip{Xf}{Xg}$, since in this space $\hat H$ turns out to be self-adjoint.

\vspace{2mm}

{\bf Remarks:--} (1) In \cite{ban}, the reason why a procedure of symmetrization was used, was to cancel certain poles in some relevant wave-functions. Changing scalar product may produce the same effect: for instance, if $\hat x$ is a pole of the third order for a wave function $\Psi(x)$ of our (one-dimensional) system, it is sufficient to modify the scalar product with a measure containing $(x-\hat x)^6\,e^{-x^2}$, for instance. In this case, in fact, $\int_{\Bbb R}|\Psi(x)|^2\,(x-\hat x)^6\,e^{-x^2}\,dx$ can be finite.

(2) There exists a simple situation in which it is possible to check that the choice above of $\Hil_T$ is, in a certain sense, {\em
optimal}: let $T$ be unbounded with bounded $T^{-1}$. Of course, this does not appear to be the case for the Calogero model but may be true for other physical
systems. Let us consider a different Hilbert space, $\Hil_\alpha$, with scalar product $\ip{f}{g}_\alpha:=\ip{T^{\alpha}f}{T^{\alpha}g}$,
$f,g\in D(T^\alpha)$. Because of (\ref{34}), the eigenvectors of $\tilde H_D$ are necessarily of the form $T\Phi_{n_1,n_2}$, which belongs to
$\Hil_\alpha$ if $\alpha\leq-1$. Due to the uniqueness of the set $\F_\Psi$, we must have  $\Psi_{n_1,n_2}=T^{-(1+2\alpha)}\Phi_{n_1,n_2}$ and
$\Psi_{n_1,n_2}\in\Hil_\alpha$ if $\alpha\geq-1$. In fact, with this choice,
$\ip{\Psi_{n_1,n_2}}{\varphi_{m_1,m_2}}_\alpha=\delta_{n_1,m_1}\delta_{n_2,m_2}$. Hence, the only choice which allows $\varphi_{n_1,n_2},
\Psi_{n_1,n_2}\in\Hil_\alpha$, is $\alpha=-1$: we recover our Hilbert space $\Hil_T$.

\section{Conclusions}

We have shown how the QHO could be treated in a weighted $\Lc^2$ space and how this choice produces an interesting functional structure, where
self-adjoint operators are replaced by pseudo-hermitian operators and where o.n. bases are replaced by biorthogonal sets. This procedure can be
interesting to {\em enlarge} the Hilbert space, so to include more functions, not necessarily asymptotically decaying, into the game.

Using the $D_2$ type Calogero model we have also shown how the choice of the Hilbert space is  suggested by the hamiltonian of the system, and
from possible existing intertwining operators between this hamiltonian and some related self-adjoint operator. In particular, we have used this
idea to deduce a convenient Hilbert space to be used in the analysis of this model. This conclusion reflects very much those  of our older
analysis concerning the quantum dynamics of quantum systems at an algebraic level, \cite{bagrmp}, where we have shown how the hamiltonian of a
system determines, by itself, the algebraic and the topological settings to be used for the analysis of the system.

Of course, it would be interesting to deduce the explicit form of $D(T^{-1})$ and to prove if $T^{-1}$ is closable. From this point of view, therefore, the results of Section \ref{sectbtnfc} require some extra  {\em mathematical} effort. This is work in progress.

\section*{Acknowledgements}
   I would like to thank Prof. Ragnisco and Dr. Rigoni for useful discussions at a first stage of this paper. I also acknowledge financial support from MIUR.

\renewcommand{\theequation}{A.\arabic{equation}}

\section*{Appendix:  Back to the QHO: $\F_\varphi$ and $\F_\Psi$ are bases for $\Hil_\pi$}

We have already seen that  $\F_\varphi$ and $\F_\Psi$ are biorthogonal and complete in $\Hil_\pi$. However, this does not automatically implies
that they are also bases for $\Hil_\pi$, \cite{heil,bagnew}, since there could exist vectors in $\Hil_\pi$ which cannot be expanded in terms of
the vectors of these sets. However, this is not the case in our model. In fact, let us define a new set $\E$ of functions
$e_{n_1,n_2}(x_1,x_2):=e^{-\frac{1}{2}\pi(x_1,x_2)}\varphi_{n_1,n_2}(x_1,x_2)$, with $\varphi_{n_1,n_2}(x_1,x_2)$ as in (\ref{extra}). It is
easy to check that $\E$ is an o.n. basis for $\Hil_\pi$. It is also clear that each $e_{n_1,n_2}$ belongs to the domain of $T$ and $T^{-1}$,
$e_{n_1,n_2}\in D(T)\cap D(T^{-1})$. Here $T$ is the following multiplication operator: $(Tf)(x_1,x_2):=e^{\frac{1}{2}\pi(x_1,x_2)}f(x_1,x_2)$,
for each $f\in D(T)$. Hence, since $Te_{n_1,n_2}(x_1,x_2)=\varphi_{n_1,n_2}(x_1,x_2)$ and, for the choice of $\pi(x_1,x_2)$ considered in
Section \ref{sec3a}, $T^{-1}e_{n_1,n_2}(x_1,x_2)=\Psi_{n_1,n_2}(x_1,x_2)$, with $\Psi_{n_1,n_2}$ given in (\ref{extra2}), from \cite{bagnew}
follows our claim: $\F_\varphi$ and $\F_\Psi$ are bases for $\Hil_\pi$.


\begin{thebibliography}{99}
\bibitem{bagrep} F. Bagarello, {\em Pseudo-bosons, so far}, Rep. Math. Phys., {\bf 68}, No. 2, 175-210 (2011)


\bibitem{bagnlpb} F. Bagarello, {\em Non linear pseudo-bosons}, J. Math. Phys.,  {\bf 52}, 063521, (2011)

\bibitem{bit2011} F. Bagarello, A. Inoue, C.Trapani, {\em Weak commutation relations of unbounded operators and applications}, J. Math. Phys.  {\bf 52}, 113508 (2011)


\bibitem{ban} P. Banerjee and B. Basu-Mallick, {\em Exact solution of $D_N$ type quantum Calogero model through a mapping to free harmonic oscillator}, J. Math. Phys., {\bf  52}, 052106 (2011)
\bibitem{nishino} A. Nishino, H. Ujino§ and M. Wadati, {\em Symmetric Fock space and orthogonal symmetric polynomials associated with the Calogero model}, Chaos, Solitons \& Fractals DOI:10.1016/S0960-0779(98)00138-6

\bibitem{bagpb1} F. Bagarello {\em Pseudo-bosons, Riesz bases and coherent states}, J. Math. Phys., {\bf 50}, DOI:10.1063/1.3300804, 023531 (2010) (10pg)


\bibitem{kolfom} A. Kolmogorov and S. Fomine, {\em El\'ements de la th\'eorie des fonctions et de l'analyse fonctionelle}, Mir (1973)

\bibitem{mosta} A. Mostafazadeh, {\em Pseudo-Hermitian representation of Quantum Mechanics}, Int. J. Geom. Methods Mod. Phys. {\bf 7}, 1191-1306 (2010)

\bibitem{bender} C. Bender, {\em Making Sense of Non-Hermitian Hamiltonians}, Rep. Progr.  Phys., {\bf 70},  947-1018 (2007)

\bibitem{bagrmp} F. Bagarello {\em Algebras of unbounded operators and physical
applications: a survey},  Reviews in Math. Phys, {\bf 19}, No. 3, 231-272 (2007)

\bibitem{heil} C. Heil, {\em A basis theory primer: expanded edition}, Springer, New York, (2010)

\bibitem{bagnew}  F. Bagarello, {\em More mathematics for pseudo-bosons}, Ann. of Phys., submitted.


\end{thebibliography}
\end{document}